\documentclass[twocolumn,showpacs,preprintnumbers,pra]{revtex4}
\usepackage{graphicx}
\usepackage{dcolumn}
\usepackage{bm}
\usepackage{color}

\begin{document}

\title{Suppression of Phase Decoherence in a Single Atomic Qubit}
\author{Chih-Sung Chuu$^1$}
\author{Chuanwei Zhang$^2$}
\thanks{Correspondence should be addressed to cwzhang@wsu.edu}
\affiliation{$^1$Edward L. Ginzton Laboratory, Stanford University, Stanford, California
94305, USA\\
$^2$Department of Physics and Astronomy, Washington State University,
Pullman, Washington 99164, USA }

\begin{abstract}
We study the suppression of noise-induced phase decoherence in a single
atomic qubit by employing pulse sequences. The atomic qubit is composed of a
single neutral atom in a far-detuned optical dipole trap and the phase
decoherence may originate from the laser intensity and beam pointing
fluctuations as well as magnetic field fluctuations. We show that suitable
pulse sequences may prolongate the qubit coherence time substantially as
compared with the conventional spin echo pulse.
\end{abstract}

\maketitle

Suppressing decoherence in a quantum system is of great importance for
quantum information processing as well as high precision spectroscopy. The
fault-tolerance quantum computation requires the decoherence rate to be
below a threshold level~\cite{QI_book}. Low decoherence is also demanded to
store quantum information in a quantum memory \cite{Lukin_QM2003,
Polzik_QM2004,QI_QM1,Pan_QM2009,Kuzmich_QM2009,Monroe_PRL2009}. For high precision
spectroscopy, suppressing decoherence prolongs the measurement time and thus
increases the precision of the measurement. In view of achieving long
coherence times, many quantum information processing protocols \cite%
{Zoller_QI1995,Mabuchi_QI1997,Zoller_QI1997,Deutsch_QI1999,
Zoller_QI1999,Molmer_QI1999,Polzik_QI1999,Zoller_QI2000,Lukin_QI2000,
Polzik_QI2000,QI_DLCZ,Monroe_2004,Monroe_JOSAB2007,Monroe_2008} and high precision
measurements \cite%
{Chu_PM1992,HighPrecisionSpec,Kasevich_PM1997,Chu_PM1999,Katori_PM2005,
Ye_PM2005,Ye_PRL2007} have thus employed the long-lived internal states of ions or
neutral atoms.

However, a quantum system cannot be completely isolated from the
environment, leading to unavoidable decoherence for quantum states.
Therefore a critical question is how to suppress the decoherence to a
desired level for various applications. In this paper, we consider the
suppression of the phase decoherence in an atomic qubit which is composed of
a single neutral atom confined in a far-detuned optical dipole trap. The
qubit is defined by two hyperfine states of the atom. This system is an
excellent candidate for quantum computation because it is well isolated from
the environment and is also easy to be exploited for storing and processing
quantum information. In this system, there are two important types of
decoherence mechanisms. The first is the spin relaxation, originating from
the inelastic Raman scattering (IRS) of photons from the trapping laser or
the spin exchange collision in hyperfine manifolds \cite{Dipole_Trap_Review}%
. The corresponding decoherence time is known as $T_{1}$. The second type of
decoherence mechanism is the fluctuations of laser and magnetic field
intensities as well as trap positions, which can modulate the energy
splitting between two qubit states and thus lead to phase decoherence of the
qubit and loss of quantum information. This type of decoherence is known as
dephasing with a decoherence time $T_{2}$. In far-detuned optical traps, the
IRS is greatly suppressed because of the large detunings \cite{Heinzen}. As
a result, $T_{1} $ can be very long and $T_{2}\ll T_{1}$. The suppression of
phase decoherence is hence most relevant to the quantum information
processing and quantum measurements in a far-detuned optical trap.

How to suppress the phase decoherence in various quantum systems has
attracted much attention both theoretically and experimentally. For many
years in the field of nuclear magnetic resonance, applications of external
pulse sequences have been investigated in order to refocus the phase
diffusion or decouple the qubit from the environment~\cite{NMR}. Some of
these techniques have been applied to superconducting qubits where
significant enhancement of decoherence time has been observed~\cite{SSQubit}%
. Recently, composite pulses have been employed onto an ensemble of atomic
qubits \cite{Davidson_Composite,Bollinger,Porto}. Hahn's spin echo (SE)
sequence \cite{SE} has also been implemented for an atomic ensemble in an
optical dipole trap \cite{Davidson_Spin_Echo,Meschede_decoherence_analysis}
to enhance phase coherence time. Here, we investigate the performance of
more elaborate pulse sequences on suppressing the noise-induced phase
decoherence of the single atomic qubit. We find that multi-pulse sequences
outperform the conventional SE sequence by orders of magnitude.

Common origins of decoherence for a single atomic qubit in an optical dipole
trap are laser intensity fluctuations, beam pointing fluctuations, and
magnetic field fluctuations.

(i) \textit{Laser intensity fluctuations}. In a single atomic qubit,
magnetic Zeeman sublevels are often exploited as the qubit basis \cite%
{Meschede_single_atom,Grangier_single_atom,Weinfurter_single_atom}. For
example, we can define a qubit using $\left\vert \downarrow \right\rangle
=|5S_{1/2},F_{1}=1,m_{F_{1}}=0\rangle $ and $\left\vert \uparrow
\right\rangle =|5S_{1/2},F_{2}=2,m_{F_{2}}=0\rangle $ states of $^{87}$Rb
atoms. The energy splitting $E(r,t)$ of the qubit in an optical dipole trap is
related to the intensity of the trapping laser $I(r,t)$ through
\begin{equation}  \label{splitting}
E\left(r, t\right) =E_{H}+\frac{\pi c^{2}\Gamma }{2\omega _{0}^{3}}\left(
\frac{1}{\Delta _{F_{2}}^{^{\prime }}}-\frac{1}{\Delta _{F_{1}}^{^{\prime }}}%
\right) I\left(r, t\right) ,  \label{energy_splitting}
\end{equation}%
where $E_{H}$ is the hyperfine splitting between two qubit states without
the laser field, $\Gamma $ is the natural linewidth, $\omega _{0}$ is the
atomic transition frequency, and $1/\Delta _{F}^{^{\prime }}=(2+\alpha
g_{F}m_{F})/\Delta _{2,F}+(1-\alpha g_{F}m_{F})/\Delta _{1,F}$. The quantity
$\alpha =\left\{ 1,0,-1\right\} $ denotes the polarization of the trapping
laser, and $\Delta _{2,F}$ ($\Delta _{1,F}$) is the detuning with respect to
the atomic transition $\left\{ 5S_{1/2},F\right\} \rightarrow 5P_{3/2}$ ($%
5P_{1/2}$). The laser intensity fluctuations, $I(t)=I_0[1+\beta (t)]$, thus
result in temporal fluctuation of the energy splitting $\delta E(t) = E_L
\beta (t)$, which in turn induces dephasing.

(ii) \textit{Beam pointing fluctuations}. The spatial dependence of $I(r,t)$
in Eq.~\ref{splitting} for a focused Gaussian-beam is given by $I(r)=I_{0}\
\mathrm{exp}(-r^{2}/2w_{0}^{2})$, where $r$ is the position of the atom with
respect to the trap center, $I_{0}$ is the peak intensity, and $w_{0}$ is
the beam waist. The beam pointing fluctuations may originate from the air
turbulence or mechanical vibration of the mirrors and lenses along the beam
path. As a consequence, the position of the trap center $\gamma (t)$
fluctuates with time, leading to $r(t)=r_{0}-\gamma (t)$, where $r_{0}$ is
the actual position of the atom. In experiments, the position fluctuations $%
\gamma (t)$ may be suppressed to the order of 10 nm for a typical beam waist
of $\sim 5\ \mu $m. Since $r\ll w_{0}$, we can approximate the trapping
potential by a harmonic trap. For an atom in the ground state of the trap,
$\gamma (t)$ is much smaller than the atom's average position $\bar{r}_{0}\sim
\sqrt{\hbar /m\omega }\sim 100$~nm for a typical trapping frequency $\omega
\sim 2\pi \times 10$ kHz. In addition, the beam pointing fluctuations are
only significant for frequency $\bar{\omega}$ below tens of Hz \cite%
{Thomas_noise}. The atom thus follows the vibration of the trap adiabatically
because the moving velocity of the trap $v_{t}\sim \gamma (t)\bar{\omega}$
is much smaller than the atom's velocity $v_{a}\sim \bar{r}_{0}\omega $,
leading to the satisfaction of the adiabatic condition
\begin{equation}
\hbar v_{t}\ll \frac{E_{g}^{2}}{\left\vert \left\langle \psi _{g}\right\vert
\frac{\partial H}{\partial \gamma (t)}\left\vert \psi _{e}\right\rangle
\right\vert }.
\end{equation}%
Here $E_{g}$ is the energy gap between the ground state $\psi _{g}$ and the
excited states $\psi _{e}$ of the harmonic trap, $H=p^{2}/2m+m\omega
^{2}r^{2}(t)$ is the Hamiltonian of the system. As a result, the low-frequency 
beam pointing fluctuations do not induce dephasing in the atom qubit
since the atom feels the same trapping potential even if the trap center fluctuates.
Moreover, the high-frequency part of the beam pointing fluctuations
only leads to negligible dephasing for the atom qubit because of its low
magnitude \cite{Thomas_noise}. The dephasing associated with the beam pointing fluctuations is
thus not significant. The heating resulted from the beam pointing
fluctuations, on the other hand, may induce dephasing but it is negligible
within the time scale of the trap lifetime \cite{Thomas_heating}.

(iii) \textit{Magnetic field fluctuations}. In the presence of a weak
magnetic field $B_{z}$, the energy levels of the atom split linearly
according to $E_{B}=m_{F}g_{F}\mu _{B}B_{z}\propto m_{F}I_{B}$, where $I_{B}$
is the current of the Helmholtz coil used for generating the magnetic field.
Therefore the classical noise of the current source $\delta I_{B}(t)$ may
give rise to fluctuation of the energy splitting of the qubit, namely, $%
\delta E(t)\propto (m_{F_{2}}-m_{F_{1}})\delta I_{B}(t)$. In
experiments, however, this can be avoided by making use of clock states, such
as the superposition state of $|5S_{1/2},F_{1}=1,m_{F_{1}}=-1\rangle $ and $%
|5S_{1/2},F_{1} =2,m_{F_{1}}=1\rangle $ or the $m_{F}=0$ Zeeman sublevels in
two hyperfine states \cite{Davidson_Spin_Echo,Meschede_decoherence_analysis,
Clock_Cornell,Clock_Hansch,Monroe_PRA2005,Monroe_PRA2007}, for the qubit states. 
For example, the latter has been employed to
achieve a coherent time exceeding 15 minutes for an atomic clock reported in
Ref.~\cite{Clock_15min}. As a result, the energy splitting of the qubit is
unaffected by the temporal fluctuation of the magnetic field.

To study the dephasing, we consider the following Hamiltonian for a single
atomic qubit,
\begin{equation}
\hat{H}=\frac{1}{2}\left[ E_{0}+\epsilon \left( t\right) \right] \hat{\sigma
_{z}},  \label{hamiltonian}
\end{equation}%
where $\epsilon \left( t\right) $ represents the temporal fluctuation of the
energy splitting with respect to the average splitting $E_{0}$. We first
assume that one noise source is dominant. Later on, we will discuss the case
in which one needs to take into account multiple noise sources.

In the experiments for studying the decoherence time, one usually prepares
the qubit first in the eigenstate of $\hat{\sigma _{z}}$, e.g., $\left\vert
\uparrow \right\rangle $, by means of optical pumping. Subsequently, a
microwave or two-photon Raman $\pi /2$-pulse initializes the qubit in its
superposition state $\left\vert \psi \left( 0\right) \right\rangle
=(\left\vert \uparrow \right\rangle +\left\vert \downarrow \right\rangle )/%
\sqrt{2}$ at $t=0$ with the off-diagonal density matrix element being $\rho
_{\uparrow \downarrow }\left( 0\right) =1/2$. Then, after a freely evolving
time $t$ in a free induction decay (FID) experiment, the qubit state becomes
\begin{equation}
\left\vert \psi \left( t\right) \right\rangle =\frac{1}{\sqrt{2}}\left(
e^{i\phi _{\uparrow }/2}\left\vert \uparrow \right\rangle +e^{i\phi
_{\downarrow }/2}\left\vert \downarrow \right\rangle \right) ,
\label{qubit_state}
\end{equation}%
where $\phi _{\uparrow }=-\phi _{\downarrow }=-\int_{0}^{t}\epsilon \left(
t^{\prime }\right) dt^{\prime }/2$ in a rotating reference frame. The qubit
state thus accumulates a phase $\Delta \phi =\phi _{\uparrow }-\phi
_{\downarrow }$ during the free evolution of time $t$ and the off-diagonal
density matrix element evolves according to
\begin{equation}
\rho _{\uparrow \downarrow }\left( t\right) =\rho _{\uparrow \downarrow
}\left( 0\right) \left\langle e^{-i\Delta \phi (t)}\right\rangle ,
\label{density_matrix}
\end{equation}%
where $\left\langle \cdots \right\rangle $ denotes averaging over an
ensemble of identical systems. For fluctuations whose statistics is
stationary, the ensemble average is equivalent to the time average.

To characterize the dephasing for a qubit, we define the decoherence
function $W(t)$ to be
\begin{equation}  \label{decoherence_function}
W\left(t\right) \equiv \frac{\left| \rho_{\uparrow\downarrow}
\left(t\right)\right|}{\left| \rho_{\uparrow\downarrow} \left(0\right)\right|%
}.
\end{equation}
Thus, $W(t)=1$ if there is no dephasing and $W(t)<1$ if there is dephasing.
For a FID experiment, it can then be shown that \cite{Cywinski}
\begin{equation}  \label{FID_decoherence}
W^{\mathrm{FID}}(t) = \mathrm{exp} \left( - \int_0^{\infty} \frac{d \omega}{%
\pi} S(\omega) \frac{2 \mathrm{sin}^2 \frac{\omega t}{2}}{\omega^2} \right),
\end{equation}
where $S(\omega)$ is the power spectrum or the first spectral density of the
noise, i.e., the Fourier transform of the correlation function $S(t)=\langle
\epsilon (t) \epsilon (t+\tau) \rangle$ of the noise. The decoherence
function is not necessary a Gaussian function, but one can still define the
decoherence time $T_2$ to be $W(T_2) = 1/e$ for convenience.

Now, we consider simultaneous presence of multiple noise sources $\beta
_{i}\left( t\right) $. In this case, the correlation function is given by $%
S\left( t_{1}-t_{2}\right) =\langle \sum_{i}\epsilon _{i}\left( t_{1}\right)
\sum_{j}\epsilon _{j}\left( t_{2}\right) \rangle $. If the noise sources are
uncorrelated, i.e., $\langle \epsilon _{i}(t_{1})\epsilon _{j}(t_{2})\rangle
=\delta _{ij}S_{i}\left( t_{1}-t_{2}\right) $, the correlation function can
be reduced to $S(t_{1}-t_{2})=\sum_{i}S_{i}(t_{1}-t_{2})$. The power
spectrum of the noise is then given by the summation of individual power
spectrum, $S\left( \omega \right) =\int_{-\infty }^{\infty }e^{i\omega
t}S\left( t\right) dt=\sum_{i}S_{i}\left( \omega \right) $, where $%
S_{i}\left( \omega \right) =\int_{-\infty }^{\infty }e^{i\omega
t}S_{i}\left( t\right) dt$. Accordingly, the decoherence function is the
product of each decoherence function, $W(t)=\prod_{i}W_{i}(t)$. We see that
the decoherence is dominated by the noise source with shorter dephasing
time.

\begin{figure}[t]
\includegraphics[width=0.9\linewidth]{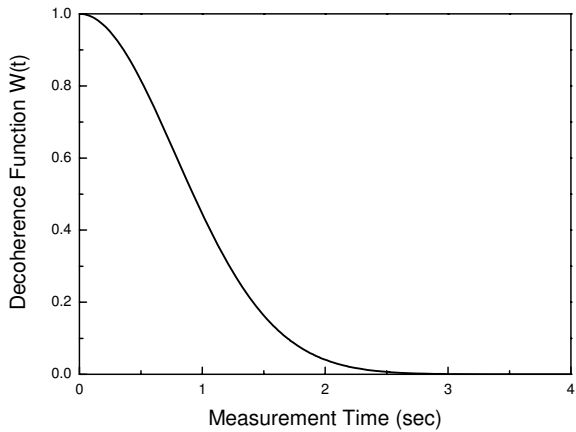}
\caption{Decoherence function for a simulated FID experiment in the
presence of intensity fluctuation of the trapping laser.}
\label{fig:1}
\end{figure}

Fig.~\ref{fig:1} shows the decoherence function for a single atomic qubit in
a simulated FID experiment. The decoherence time is found to be $T_{2}\sim 1$
~sec. The two qubit states are $\left\vert \downarrow \right\rangle =
|5S_{1/2},F_{1}=1,m_{F_{1}}=0\rangle$ and $\left\vert \uparrow \right\rangle
=|5S_{1/2},F_{2}=2,m_{F_{2}}=0 \rangle $ states of $^{87}$Rb atom. The atom
is trapped at the bottom of an optical dipole trap that is generated by a
YAG laser with a trap depth of $\sim \ 500\ \mu $K. The
only relevant classical noise taken into account here is the intensity
fluctuation of the trapping laser. The power spectrum is adopted from Ref.~%
\cite{YAG}, which can be approximated by $S(f)/E_L^2 = 10^{-8.5}f^{-5/3}$ Hz$%
^{-1}$ for frequencies below 1~kHz. As the longest trap lifetime reported
thus far is $\sim 400$ sec, we choose the infrared cutoff frequency to be $%
\omega _{ir}\sim 0.016$~s$^{-1}$. We note that the decoherence time depends 
strongly on the characteristics of the power spectrum. For the power spectrum
given in Ref.~\cite{Markus_thesis}, we obtain $T_{2}\sim 20$~ms for the
same trap configuration.

Dephasing in a single atomic qubit may be reversed by applying a sequence of
$\pi $ pulses. The simplest case is a SE sequence in which one applies a
microwave or two-photon Raman $\pi $ pulse at halftime $\tau $ of the free
evolution. By doing this, one can partially cancel the dephasing due to
low-frequency ($<1/\tau $) noise. However, SE becomes less effective when
high-frequency noise is present. Furthermore, the imperfection of the $\pi $
pulse inherently introduces additional phase diffusion onto the qubit state
(for example, one applies a \textquotedblleft $\pi +\delta $%
\textquotedblright\ pulse with $\delta <\pi $ instead of a $\pi $ pulse).
Accordingly, multi-pulse sequences may be a better choice for suppressing the 
dephasing more effectively as well as compensating the phase
error of the $\pi $ pulses.

We consider a general pulse sequence that is composed of $n$ instantaneous $%
\pi $ pulses at time $t_{1},t_{2},\ldots ,t_{n}\in \left[ 0,t\right] $. The $%
\pi $ pulse rotates the qubit state about the $x$-axis, therefore the qubit
state after the application of the pulse sequence evolves as
\begin{eqnarray}
\left\vert \psi \left( t\right) \right\rangle &=&e^{-i\int_{t_{n}}^{t}\hat{H}%
(t^{\prime })dt^{\prime }}(-i\hat{\sigma}_{x})\cdots
e^{-i\int_{t_{1}}^{t_{2}}\hat{H}(t^{\prime })dt^{\prime }}  \nonumber
\label{state_after_pulses} \\
&&(-i\hat{\sigma}_{x})e^{-i\int_{0}^{t_{1}}\hat{H}(t^{\prime })dt^{\prime
}}\left\vert \psi \left( 0\right) \right\rangle .
\end{eqnarray}%
The decoherence function defined in Eq.~\ref{decoherence_function} can then
be shown to be \cite{Cywinski}
\begin{equation}
W(t)=\mathrm{exp}\left( -\int_{0}^{\infty }\frac{d\omega }{\pi }S(\omega )%
\frac{F(\omega t)}{\omega ^{2}}\right) ,
\label{decoherence_function_after_pulses}
\end{equation}%
where $F(\omega t)=\frac{1}{2}|\sum_{k=0}^{n}(-1)^{k}(e^{i\omega
t_{k+1}}-e^{i\omega t_{k}})|^{2}$ corresponds to a certain pulse sequence
which has a specific set of $t_{k}$ with $t_{0}=0$ and $t_{n+1}=t$. In the
following, we focus on the performance of various pulse sequences listed
below.
\begin{figure}[t]
\includegraphics[width=0.9\linewidth]{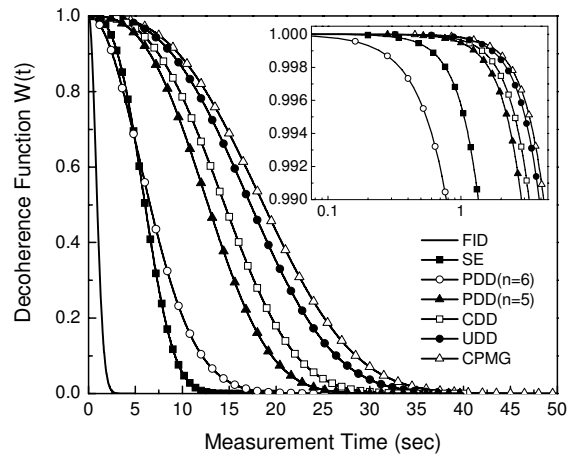}
\caption{Decoherence functions with applications of SE, PDD, CDD, UDD, and
CPMG pulse sequences. FID is also shown for comparison. The inset shows the
short-time performance for various pulse sequences.}
\label{fig:2}
\end{figure}

\textit{\textrm{SE} Pulse Sequence}. SE is an efficient technique to reverse
the low-frequency dephasing which exists prior to the application of the $%
\pi $ pulse. The pulse sequence comprises a single $\pi $ pulse at $%
t_{k}=t/2 $ ($n=1$) with $F(\omega t)=$ $8\ \mathrm{sin}^{4}(\omega t/4)$.

\textit{Carr-Purcell-Meiboom-Gill \textrm{(CPMG)} Pulse Sequence}. CPMG is
the $N$ times repetition of SE sequence \cite{CP,MG}. For CPMG, we have $%
t_{k}=(k-1/2)\ t/n$ and $F(\omega t)=$ $8\ \mathrm{sin}^{4}(\omega
t/4n)G(\omega t)\mathrm{cos}^{-2}(\omega t/2n)$, where $G(\omega t)=\mathrm{%
sin}^{2}(\omega t/2n)$ for even $n$ and $G(\omega t)=\mathrm{cos}^{2}(\omega
t/2n)$ for odd $n$.

\textit{Periodic Dynamical Decoupling \textrm{(PDD)} Pulse Sequence}.
Dynamical decoupling (DD) sequences are designed to decouple the qubit from
the influence of environment. For PDD, the $n$ pulses are equally
distributed over the entire measurement time: $t_{k}=kt/(n+1)$ and $F(\omega
t)=$ $2\ \mathrm{tan}^{2}[\omega t/(2n+2)][1-G(\omega t)]$. A property of
PDD is that only the odd order of the sequence can suppress the
low-frequency noise ($\omega <2/t$) \cite{Cywinski}.

\textit{Concatenated Dynamical Decoupling \textrm{(CDD)} Pulse Sequence}.
CDD is a concatenated DD sequence~\cite{CDD}. The \textit{l}-th order of the
pulse sequence $\mathrm{CDD}_{l}(t)$ is defined as $\mathrm{CDD}%
_{l-1}(t/2)\rightarrow \Pi \rightarrow \mathrm{CDD}_{l-1}(t/2)\ \mathrm{for\
odd\ \mathit{l}}$ and $\mathrm{CDD}_{l-1}(t/2)\rightarrow \mathrm{CDD}%
_{l-1}(t/2)\ \mathrm{for\ even\ \mathit{l}}$, where $\Pi $ refers to an
instantaneous $\pi $ pulse and $\mathrm{CDD}_{0}(t)$ denotes free evolution
for duration $t$. As a result, $F(\omega t)=2^{2l+1}\mathrm{sin}^{2}(\omega
t/2^{2l+1})\prod_{1}^{l}\mathrm{sin}(\omega t/2^{k+1})$ with $l\approx {\log
}_{2}n$.

\textit{Uhrig Dynamical Decoupling \textrm{(UDD)} Pulse Sequence}.
Originally proposed by Uhrig \cite{Uhrig}, UDD was later shown to be an
optimal DD sequence when the delay times between pulses are sufficiently
short \cite{UDD_DasSarma}. For UDD, the sequence is defined as $t_{k}=%
\mathrm{sin}^{2}[\pi k/(2n+2)]t$ and $F(\omega t)=\frac{1}{2}%
|\sum_{-n-1}^{n}(-1)^{k}\mathrm{exp}\left\{ \mathrm{cos}[\pi k/(n+1)]\omega
t/2\right\} |^{2}$.

\begin{figure}[t]
\includegraphics[width=0.9\linewidth]{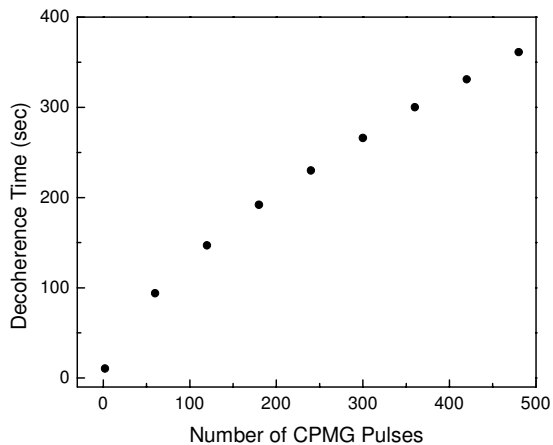}
\caption{Decoherence time as a function of number of CPMG pulses applied to
a single atomic qubit.}
\label{fig:3}
\end{figure}

The decoherence functions with the applications of various pulse sequences as
well as free evolution (FID) are shown in Fig.~\ref{fig:2}. The number of
pulses used during the measurement time is $n=6$. For PDD, 5-pulse sequence
is also shown. One can see that the even order ($n=6$) of PDD sequence is
less effective than the odd order ($n=5$) sequence. Due to the presence of a
substantial portion of low-frequency noise in the power spectrum, SE
sequence already exhibits a pronounced prolongation of decoherence time.
Nonetheless, multi-pulse sequences (CDD, UDD, CPMG, and odd-$n$ PDD) still
outperform SE by prolonging the decoherence time for more than a factor of
20 as compared to FID. Moreover, for short-time performance (inset of Fig. ~%
\ref{fig:2}), multi-pulse sequences are apparently more effective than SE.
This could be useful when high fidelity but not long coherence time is
preferred.

Among different multi-pulse sequences, CPMG is the most effective sequence
in terms of number of pulses. We investigate further prolongation of
decoherence time by applying more CPMG pulses. As shown in Fig.~\ref{fig:3},
the decoherence time increases approximately linearly with the number of
pulses. For 50 pulses, the decoherence time is prolonged by a factor of 100;
for 500 pulses, the decoherence time is prolonged by a factor of 350. Since
the length of a $\pi$ pulse can be as short as $\sim 10\ \mu $s, the
decoherence time is eventually limited by the lifetime of the atom in the
trap.

In summary, we have examined the performance of variety of external pulse
sequences on the suppression of phase decoherence in a single atomic qubit.
We find that, at $n=6$, pulse sequences ($n=5$ for PDD) already outperform
SE by more than a factor of 2 in terms of decoherence time. Among the pulse
sequences considered here, CPMG sequence is optimal for suppressing the
phase decoherence induced by the laser intensity fluctuations. We also show
that application of large number of CPMG pulses may achieve decoherence time
in the regime of minutes.

We thank M. G. Raizen and L. Cywinski for valuable discussion. C.Z. was
supported by the Washington State University Startup fund and the ARO.
C.-S.C. was supported by the DARPA and ARO.

\bibliography{decoherence}

\end{document}